\documentclass{PoS}
\usepackage{epsfig}

\PoS{PoS(LAT2005)075}

\title{Baryon spectroscopy with spatially improved quark 
sources\footnote{For the Bern-Graz-Regensburg (BGR) Collaboration.} }

\ShortTitle{Baryon spectroscopy with spatially improved quark sources }

\author{T.~Burch, \speaker{C.~Hagen},
        D.~Hierl, and A.~Sch\"afer\\

	Institut f\"ur Theoretische Physik\\
        Universit\"at Regensburg\\
	D-93040 Regensburg, Germany.\\

        E-mail: \email{christian.hagen@physik.uni-regensburg.de}\\
       }

\author{Christof Gattringer, L.~Ya.~Glozman\footnote{Supported by Fonds zur
    F\"orderung der Wissenschaftlichen Forschung in \"Osterreich, 
    project P16823-N08.} , and C.~B.~Lang\\
	Institut f\"ur Physik, FB Theoretische Physik\\
        Karl-Franzens-Universit\"at Graz\\
	A-8010 Graz, Austria.\\
       }

\abstract{
We study baryons on the lattice with a special focus on excited states. For 
that purpose we construct several interpolators which differ in their Dirac 
structure. These interpolators are built from Jacobi smeared quarks with 
different widths in order to allow for 
operators with improved spatial wavefunctions. 
We compute all cross correlations and use the variational method to
determine which combinations of operators have best overlap with ground and
excited states. Our approach yields promising results for the
spin-$\frac{1}{2}$ baryons:  nucleon, $\Sigma$, $\Xi$  and $\Lambda$. For the
spin-$\frac{3}{2}$ baryons, $\Delta$ and $\Omega$, we obtain results which are
consistent with results of other groups.} 

\FullConference{XXIIIrd International Symposium on Lattice Field Theory\\

                 25-30 July 2005\\

                 Trinity College, Dublin, Ireland}

\begin{document}

\section{Introduction}

Ground state spectroscopy in quenched
lattice calculations appears to be well understood. Excited state spectroscopy,
however, is still a challenging task. There are two major problems: First, one
has to improve the overlap of the interpolating fields with excited states.
Excited hadron states include also radial excitations and thus nodes in their
radial wavefunction. Allowing for such nodes is important and should be
implemented in a lattice calculation to obtain more realistic results. The
second problem is finding reliable means to disentangle ground and excited
states in the hadron spectrum. Our method of choice is a variational approach
\cite{Michael:1985ne,Luscher:1990ck}.
Some preliminary results have been presented elsewhere
\cite{Burch:2004he,Burch:2004zx,Burch:2005vn}. 
Two larger publications are in preparation.

\section{Simulation details}

We perform quenched lattice calculations using CI fermions
\cite{Gattringer:2000js,Gattringer:2000qu}. For our simulations we use two
lattices  with the same spatial volume of about $2.4$~fm. 
The lattice spacings
are $a=0.148$~fm and $a=0.119$~fm for our $16^3\times32$ and $20^3\times32$
lattices, respectively. These values are determined from the Sommer
parameter \cite{Gattringer:2001jf}. We compare our results with other ways of
setting the scale, namely the mass of the rho meson and the nucleon, and find
that they are consistent within error bars.
We are interested in baryons containing only light quarks as
well as those with strange quark content. We therefore determine the mass
parameter of the strange quark, using the pseudoscalar kaon mass as input.

\section{The Method}

\begin{figure}[t]
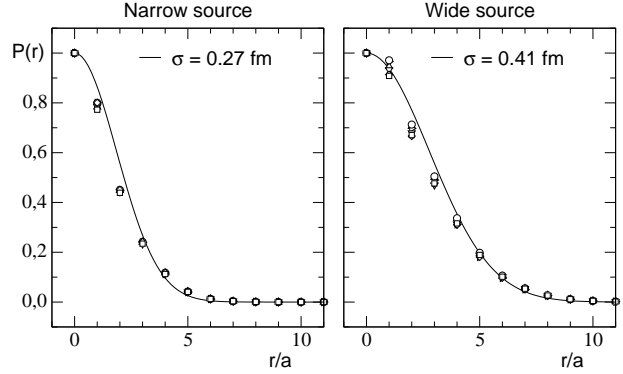

\begin{center}
\resizebox{0.54\textwidth}{!}{
\includegraphics[clip]{source_narrow.eps} \quad
\includegraphics[clip]{source_wide.eps}}
\caption{Profiles of the narrow and wide source. The symbols are our data
points and the curves are the target distributions which we approximate by the
profiles.}
\label{sources}
\end{center}
\end{figure}

In order to obtain reliable results for excited states 
it is important to have a
large basis of operators which is as complete as possible. 
One way to
increase the number of operators is to choose different Dirac structures to
combine the quark spinors. Spin and parity of the hadrons pose some
restrictions on that choice. Another way to increase the basis and to enhance
the overlap of the interpolating fields with the physical states is to improve
the spatial wavefunction of the quark sources. We use Jacobi smearing to
produce Gaussian-type gauge covariant sources with two different widths, a
narrow and a wide one (see Figure \ref{sources}). These sources can then be
combined (with a relative minus sign) to allow for a node in the radial
wavefunction of a baryon interpolator. 
The sign factors and optimal combination are {\em not} put in by hand but
chosen via  the variational method, as discussed below.
We then use these sources together with
three different Dirac structures to obtain a basis of 24 operators.

From these interpolators we construct a matrix of correlators and then apply
the variational method which has been proposed by Michael \cite{Michael:1985ne}
and later refined by L\"uscher and Wolff \cite{Luscher:1990ck}. One has to
solve a generalized eigenvalue problem. The advantage of this approach is that
the system has full freedom to choose the relative contributions of the 
different interpolators in the diagonalization. 
These weighting factors can be extracted from the components of
the eigenvectors. The mass is obtained by looking at the eigenvalues, which
in leading order behaves as
\begin{eqnarray}
\label{singleexp}
\lambda^{(k)}(t) & \propto & e^{-t \, M_k}\; .
\end{eqnarray}
From these eigenvalues we then construct effective mass plots by using
\begin{eqnarray}
M_k^{eff}\left(t+\frac{1}{2}\right) & = & 
\ln\left(\frac{\lambda^{(k)}(t)}{\lambda^{(k)}(t+1)}\right) \; .
\end{eqnarray}
Where we find a plateau in these plots we conclude that the signal of the
considered state is disentangled from higher excitations. We then fit the
eigenvalue to a single exponential according to Eq.~(\ref{singleexp}). In
the following we present the results of these fits for different baryons.
The application of our techniques to meson spectroscopy is presented in 
\cite{burch05}.

\section{Discussion of the Results}

\begin{figure}
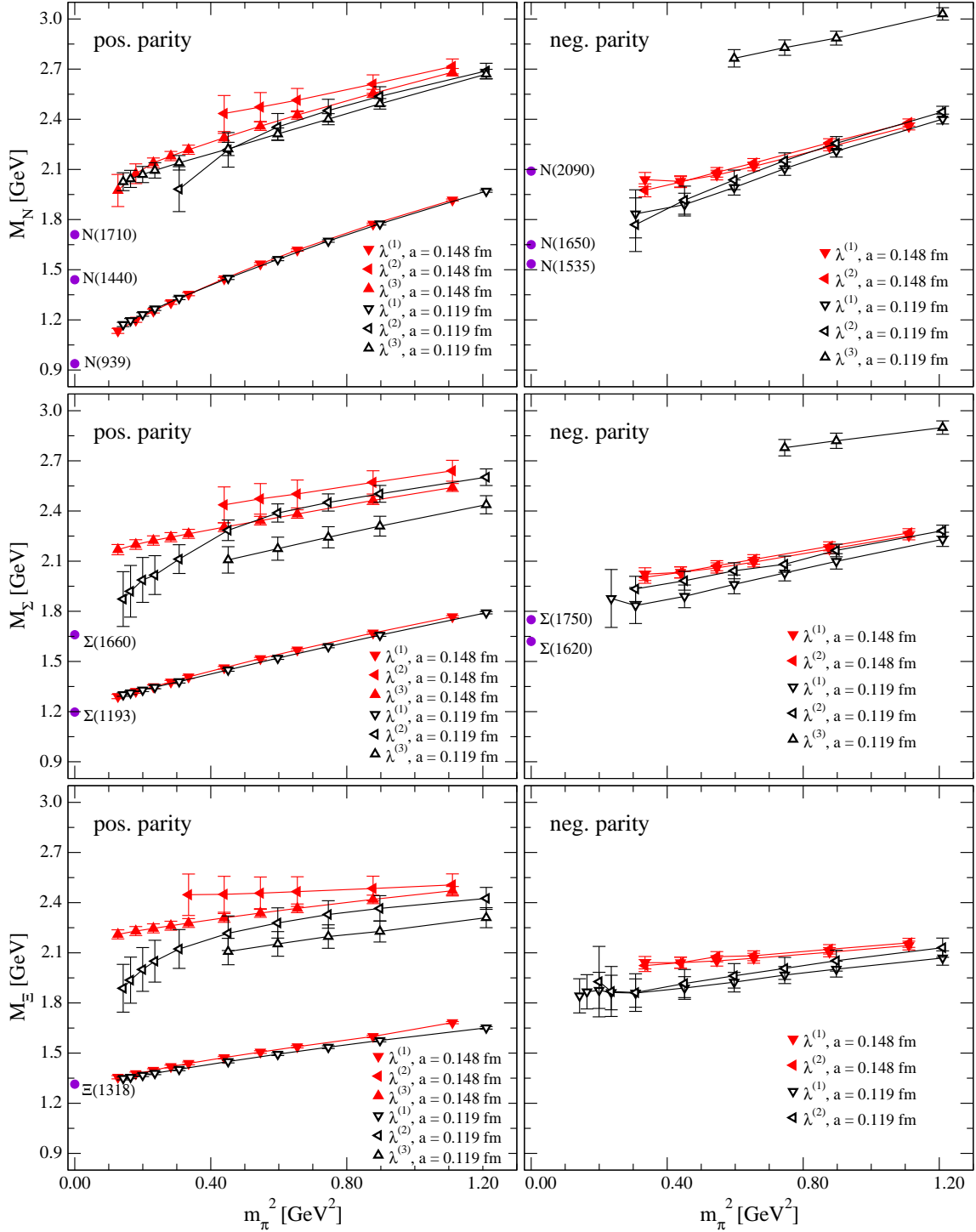

\begin{center}
\resizebox{1.00\textwidth}{!}{\includegraphics[clip]{nuc_M_vs_m_pos.eps}
\includegraphics[clip]{nuc_M_vs_m_neg.eps}}

\resizebox{1.00\textwidth}{!}{\includegraphics[clip]{sigma_M_vs_m_pos.eps}
\includegraphics[clip]{sigma_M_vs_m_neg.eps}}

\resizebox{1.00\textwidth}{!}{\includegraphics[clip]{xi_M_vs_m_pos.eps}
\includegraphics[clip]{xi_M_vs_m_neg.eps}}
\caption{Results for the nucleon, $\Sigma$, and $\Xi$
(from top to bottom) versus the pion mass squared. In each case, the left-hand
side plot shows the positive parity states while the right-hand side plot 
is for the
negative parity channel. The results obtained from the
$16^3\times32$ lattice are represented by filled symbols while those 
for $20^3\times32$ by open ones.}
\label{nuclikeresults}
\end{center}
\end{figure}

In Figure \ref{nuclikeresults} we show the results of our calculations for the
nucleon, $\Sigma$ and $\Xi$. The behavior of these baryons is very similar. 
We use interpolators of similar form, 
except for their quark content. For the nucleon
we use an interpolator with three light quarks; for $\Sigma$ and $\Xi$ we then
replace one or two of the light quarks by a strange quark with fixed 
mass parameter. Thus we can restrict ourselves to mainly discussing 
the nucleon results. The
same arguments then also hold for $\Sigma$ and $\Xi$, except that their
dependence on the pion mass squared (and thus the light quark mass) is weaker.

For the positive parity ground state nucleon, 
we find results on both lattices which
are consistent with each other and extrapolate very well to the value found by
experiment. For the first excited state in the positive parity channel, a state
which should correspond to the Roper resonance, we obtain results that are not
conclusive. For the coarser lattice we cannot identify any effective mass
plateaus at quark masses below $a\,m < 0.08$. On the finer lattice there are
indications for effective mass plateaus down to $a\,m = 0.02$, but our current
statistics does not allow for a reliable analysis (we omit data points
with $a\,m < 0.04$). However, the third eigenvalue, corresponding to the second
excitation, shows a considerably better signal. This might be understood from
the different nature of Roper and N(1710) \cite{Burch:2004he,brommel}.
In the negative parity channel, for both lattices we find two low-lying 
states which are nearly degenerate. The results on the
finer lattice are lower than for the coarse lattice and extrapolate 
reasonably well to the experimental values for N(1535), N(1650). For the 
$\Sigma$ and $\Xi$ channels we find better quality for
the Roper-like states at small quark masses. 

\begin{figure}[t]
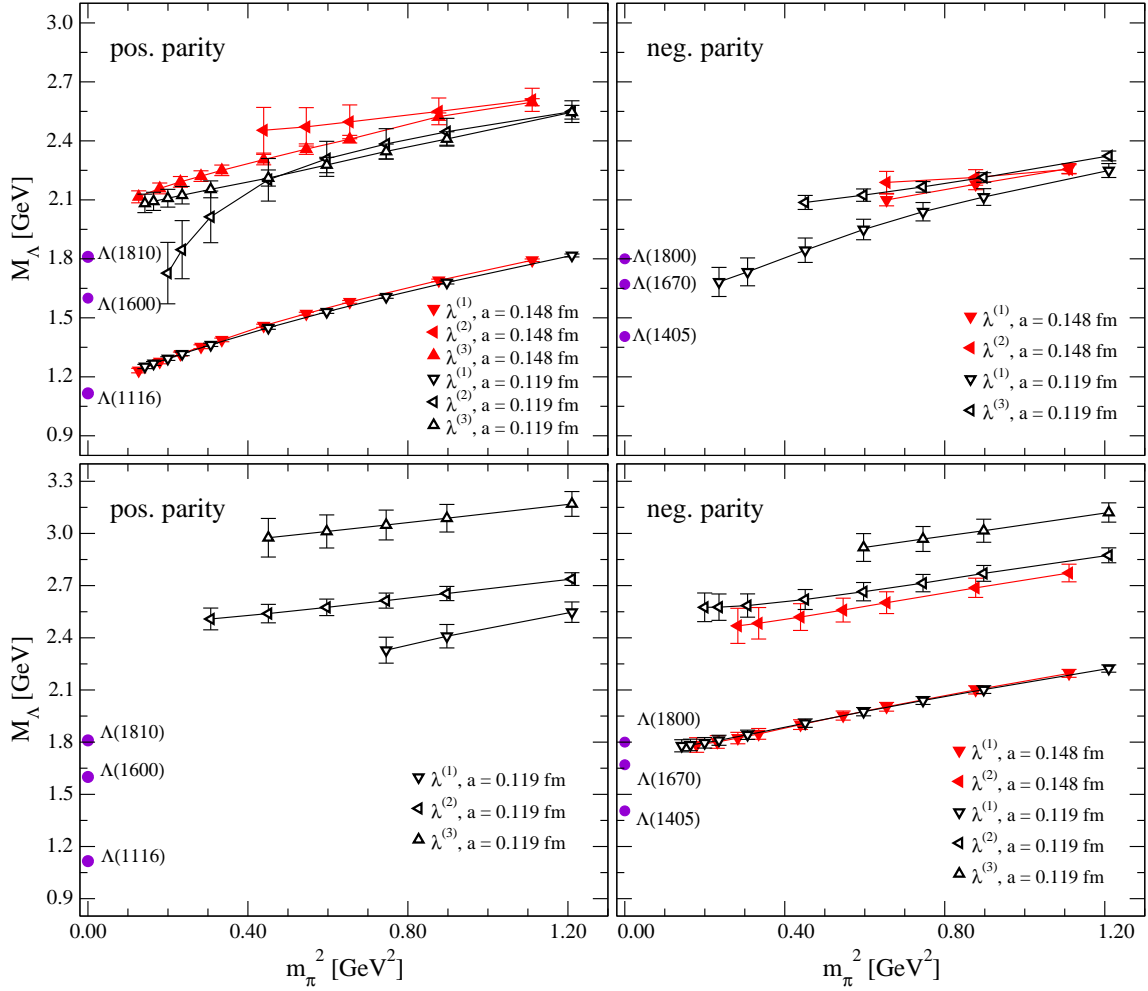

\begin{center}
\resizebox{1.00\textwidth}{!}{\includegraphics[clip]{lambda8_M_vs_m_pos.eps}
\includegraphics[clip]{lambda8_M_vs_m_neg.eps}}

\resizebox{1.00\textwidth}{!}{\includegraphics[clip]{lambda1_M_vs_m_pos.eps}
\includegraphics[clip]{lambda1_M_vs_m_neg.eps}}
\caption{Results for the $\Lambda$ flavor octet
interpolator (top) and the $\Lambda$ flavor singlet interpolator (bottom). }
\label{lambdaresults}
\end{center}
\end{figure}

For the $\Lambda$, shown in Figure \ref{lambdaresults}, we compare results
from an interpolator having mainly overlap with a flavor
octet and one which is a pure singlet. The octet shows a similar behavior as
the other octet baryons. Again, we find good
results for the positive parity ground state and a drop for the
first excited state in the positive parity channel on the finer lattice.

For the flavor singlet $\Lambda$ with positive parity we find 
effective mass plateaus only for the finer lattice and at large quark mass.
On the negative parity side, we find a strong signal for the ground state
which extrapolates essentially higher than the physical $\Lambda$(1405).
This is in agreement with previous lattice calculations.

\begin{figure}[t]
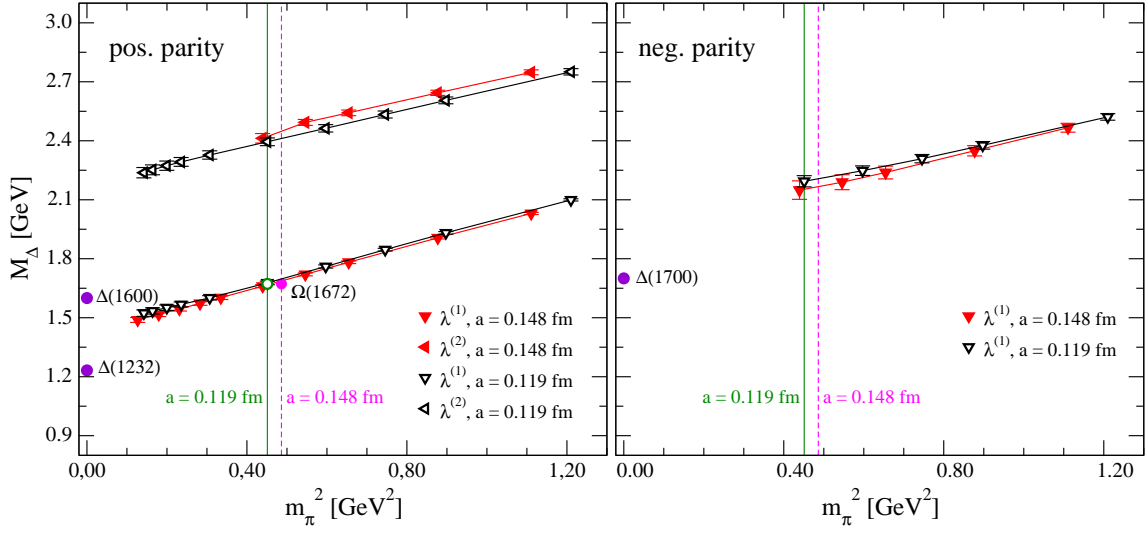

\begin{center}
\resizebox{1.00\textwidth}{!}{\includegraphics[clip]{delta_M_vs_m_pos.eps}
\includegraphics[clip]{delta_M_vs_m_neg.eps}}
\caption{Results for $\Delta$ and $\Omega$.
The vertical lines denote the pion masses squared corresponding to the
valued of the strange quark mass on our lattices (dashed for the $16^3\times32$
lattice and full for $20^3\times32$).}
\label{deltaresults}
\end{center}
\end{figure}

The results for the $\Delta$ are shown in Figure
\ref{deltaresults}. A naive extrapolation to the chiral limit 
is about 20\% too high for both parities. 
This finding is in agreement with other lattice studies. The vertical 
lines in the figure indicate the point where all three quarks have the
mass of the strange quark. The corresponding state is the $\Omega$ resonance
and we find excellent agreement with the experimental number. This indicates 
that for the $\Delta$ in full QCD, chiral dynamics, which is absent in the
quenched approximation, is important.  

\end{document}